\documentclass[aps,prb,twocolumn,groupedaddress]{revtex4-1}
\pdfoutput=1
\usepackage[pdftex]{color,graphicx}

\newcommand{\tbti}{Tb$_2$Ti$_2$O$_7$}
\newcommand{\tbsn}{Tb$_2$Sn$_2$O$_7$}
\newcommand{\tb}{Tb$^{3+}$}
\newcommand{\mub}{$\mu_B$}

\begin{document}
\title{Low temperature diffuse neutron scattering and magnetisation in single crystal \tbti}
\author{P. Bonville}
\email[To whom correspondence should be  addressed: ]{pierre.bonville@cea.fr}
\affiliation{CEA, Centre de Saclay, DSM/IRAMIS/Service de Physique de l'Etat
Condens\'e, 91191 Gif-sur-Yvette, France}
\author{A. Gukasov, I. Mirebeau, S. Petit}
\affiliation{CEA, Centre de Saclay, DSM/IRAMIS/Laboratoire L\'eon Brillouin, 91191
Gif-sur-Yvette, France}
\date{\today}

\begin{abstract}
We present an interpretation of zero field diffuse neutron scattering and of high field magnetisation data at very low temperature in the frustrated pyrochlore system \tbti. This material has antiferromagnetic exchange interactions and it is expected to have Ising character at low temperature. Contrary to expectations, it shows no magnetic ordering down to 0.05\,K, being thus labelled a ``spin liquid''. However, the ground state in \tbti\ is not a mere fluctuating moment paramagnet but, as demonstrated by very recent experiments, a state where the electronic degrees of freedom are hybridised with the phononic variables in an unconventional way. We show here that, by approximating this complex and still unraveled electron-phonon interaction by a dynamic Jahn-Teller coupling, one can account rather well for the diffuse neutron scattering and the low temperature isothermal magnetisation. We discuss the shortcomings of this picture which arise mainly from the fact that the singlet electronic mean field ground state of the model fails to reproduce the observed strong intensity of the elastic and quasi-elastic neutron scattering.
\end{abstract}


\maketitle
\section{Introduction}

The pyrochlore titanates, with formula R$_2$Ti$_2$O$_7$ where R is a rare earth, have been the subject of experimental and theoretical studies for more than a decade \cite{gardging}. The pyrochlore lattice where the R$^{3+}$ (and the Ti$^{4+}$) ions are located is formed by corner-sharing tetrahedra and leads indeed to a frustration of the exchange/dipolar interaction in some specific situations. The best known consequence thereof is the existence of ``spin-ice'' materials, like Ho$_2$Ti$_2$O$_7$ and Dy$_2$Ti$_2$O$7$ \cite{harris,ramirez}, where the rare earth moments have a strong Ising character along the tetrahedron ternary axis. The spin-ices remain in the paramagnetic phase down to the lowest attainable temperature, but the spin correlations are very strong and of the special ``two in - two out'' type, i.e. where two rare earth moments point ``outwards'' a given tetrahedron and two ``inwards``. The excitations in spin-ices have been shown to be magnetic monopole quasi-particles \cite{castel}. Another pyrochlore material which remains paramagnetic down to at least 0.05\,K is \tbti\ \cite{gardner}. The \tb\ ion is a non-Kramers ion with $J$=6 and a Land\'e factor $g_{\rm J}$=3/2. It is submitted to a trigonal symmetry crystal electric field (CEF) which lifts the 13-fold degeneracy of the $J$=6 multiplet. The peculiarity of the crystal field splitting in \tbti\ has been early recognised \cite{aleks}, as consisting of two ground magnetic doublets separated by an energy of the order of 15\,K. For a non-Kramers ion, each such doublet \{$\psi_1,\psi_2$\} has an Ising character, with a vanishing transverse matrix element of the total angular momentum: $<\psi_1 \vert {\bf J} \vert \psi_2 >$=0. The exact wave-functions of these doublets were determined \cite{gingras,mirebeau} together with the exchange contribution to the paramagnetic Curie temperature $\theta_p$, which is negative and of the order of $-$10\,K \cite{gingras,mirebeau}, indicating dominant antiferromagnetic superexchange interactions. The lack of magnetic order in \tbti\ is therefore surprising since the Ising antiferromagnet in the pyrochlore lattice is not frustrated, its ground state being the so-called ``all in / all out'' configuration.

Early neutron scattering studies \cite{gardner01,yasui} have shown that the Tb moments are very short range correlated and fluctuate down to the 0.1\,K range. Calculations of the diffuse scattering at 9\,K \cite{kao,enjalran} were successful in reproducing experimental data by taking into account the two ground doublets which are both appreciably populated at this temperature. At very low temperature, when the ground doublet alone is populated, it was recognised that Ising-like wave-functions cannot describe the physics in \tbti\ \cite{curnoe}, and that a non-zero transverse matrix element of {\bf J} is needed. In order to restore transverse spin fluctuations, a model was proposed which renormalises the low energy spin hamiltonian through quantum fluctuations between the two ground doublets $via$ virtual excitation of a third ion \cite{molavian}. It will be referred to in the following as the ``Virtual Crystal Field'' (VCF) model. It results in an effective hamiltonian containing non-Ising terms, but the ground state predicted by the VCF model cannot be the true ground state of \tbti\ since it retains magnetic ordering and thus is not a spin-liquid state \cite{molav}. In zero magnetic field, the spin liquid state seems to be fragile: magnetic ordering appears under pressure \cite{mirb_nat}, showing the great sensitivity of the spin liquid state to external stresses. It is also sensitive to the exact Tb stoichometry and may be destroyed by a slight excess in Tb content \cite{taniguchi}: small Bragg peaks are observed at ($\frac{1}{2},\frac{1}{2},\frac{1}{2}$) positions and equivalents in non-stoichiometric powder samples, whereas diffuse maxima are seen at the same positions in single crystals \cite{fennell,petit,fritsch}. Long range magnetic order develops in the presence of a magnetic field. For a field along [1$\bar1$0], a magnetically ordered phase with spin wave excitations is induced at 0.4\,K for fields of 2\,T and above \cite{rule}, and the evolution of the magnetic structure was more quantitatively explored in Ref.\onlinecite{sazonov}. For a field along [111], a similar behaviour is observed \cite{sazo2}.

Of particular significance for the understanding of the spin-liquid ground state are the numerous experimental evidences of either a dynamic symmetry lowering, by low temperature x-ray diffraction \cite{ruff}, optical measurements \cite{lummen} and inelastic neutron scattering \cite{rulebonv}, or of a strong spin-phonon coupling: the giant magnetostriction \cite{aleks,ruff2}, the low temperature divergence of the bulk modulus \cite{mamsu,mamsu88} and of the elastic constants \cite{nakanishi,luan}. Very recent high resolution neutron scattering experiments \cite{fennell2,guitt} have revealed the presence of non-standard hybridisation between an acoustic phonon branch and the first CEF transition, above an energy around 1.5\,meV, but with a vanishing low energy branch. They have also confirmed the presence of dispersive inelastic excitations around an energy of 0.2\,meV, but only in the so-called ``spin ice channel''. Finally, an anomalously strong phonon scattering at 0.3\,K has been reported \cite{li}, very likely linked with the $4f$-electron - phonon hybridisation. Hence, the ground state in \tbti\ appears to be of a quite complex and unusual kind and a full microscopic hamiltonian describing this state is still to be discovered. This hamiltonian should include a coupling between the electronic and phononic degrees of freedom, so that the ground state be of mixed vibronic type.

Previously, we proposed a model to account for the spin liquid behaviour in \tbti, assuming the presence of a static tetragonal distortion at the Tb site \cite{bonv11,bonville,petit}, i.e. of an off-diagonal crystal field term which lifts the degeneracy of the ground doublet into two CEF singlets, coupled by the exchange/dipole interaction. This model, akin to the Bleaney model for CEF singlets \cite{bleaney,cooper}, yields a strong transverse matrix element of {\bf J} and leads to a mean field paramagnetic phase, or spin liquid phase, down to the lowest temperature in a certain range of exchange integrals. This picture has been questioned in Ref.\onlinecite{gaul}, where it is claimed that no inelastic line corresponding to the splitting between singlets is observed, but the recent neutron data \cite{guitt} do show the presence of some kind of low energy inelastic excitation. On the one hand, this model satisfactorily reproduces the {\bf q}-maps of inelastic \cite{petit} and diffuse Spin-Flip and Non-Spin-Flip \cite{fennell} neutron scattering at very low temperature. On the other hand, due to the non-magnetic character of its mean field single ion ground state, this two singlet model fails to account for the strong intensity of the quasi-elastic neutron scattering \cite{petit,fennell} and for the nuclear anomaly in the specific heat \cite{yaouanc} showing the presence of a hyperfine field of 142\,T at the Tb site. Seemingly, the vibronic ground state in \tbti\ has some magnetic character.

In this work, we are interested in the interpretation of the zero-field diffuse scattering and of the isothermal magnetisation for fields applied along [111], [110] and [001] at very low temperature in \tbti. We present zero-field and in-field 3D maps of the total diffuse neutron scattering at 0.16\,K and we use previously published magnetisation data \cite{lhotel,lhotel1}. We apply a generalisation of the two singlet model to a dynamic Jahn-Teller (JT) coupling, which seems better adapted, in first approximation, to describe the low temperature state of \tbti\ since no static distortion has been evidenced. In a recent work \cite{sazo2}, we have shown that the magnetic structure induced by a field along [111] can be explained in terms of such a dynamic Jahn-Teller coupling. Here, we show that the model also satisfactorily reproduces: (i) the {\bf q}-structure of zero-field diffuse neutron scattering maps at 0.16\,K and (ii) the magnetisation curves at 0.05\,K along the 3 cubic symmetry axes. 

\section{The dynamic Jahn-Teller coupling model and the self-consistent mean field and RPA calculations}
\label{rpa}

The trigonal crystal field acting on a \tb\ ion is taken as in Ref.\onlinecite{cao09}. We emphasize that, contrary to theories describing the ground non-Kramers doublet by a pseudo-spin 1/2 and using effective exchange parameters \cite{onoda1}, we consider the whole CEF level scheme and the actual momentum {\bf J}, and we use the bare anisotropic exchange integrals ${\cal J}_{ij}$. We feel it important to recall that a non-Kramers doublet $\vert \psi_{1,2} \rangle$ presents the two properties, in the local frame:
\begin{eqnarray}\label{nonk1}
\langle \psi_1 \vert J_z \vert \psi_1 \rangle &=&-\langle \psi_2 \vert J_z \vert \psi_2 \rangle \\
\label{nonk2}
\langle \psi_2 \vert {\bf J} \vert \psi_1 \rangle &=& 0.
\end{eqnarray}
Then, considering the projection of the CEF Hamiltonian onto the subspace spanned by $\vert \psi_{1,2} \rangle$ can lead to some confusion since the z-components alone of the pseudo-spin behave as those of a true spin 1/2, but its transverse components must be chosen as quadrupolar moments \cite{onoda1}. An important consequence of (\ref{nonk2}) is that transitions between $\vert \psi_1 \rangle$ and $\vert \psi_2 \rangle$ induced by the operator {\bf J} are forbidden: there can be neither exchange/dipole induced fluctuations nor neutron quasielastic scattering within the doublet. The magneto-elastic interaction describing the coupling of the electronic variables with the local strain variables \cite{gehring} has been shown to play an important role in \tbti\ \cite{aleks,kleko}. It can be written, in the local frame, up to second order in $J_i$: 
\begin{equation}\label{hmel}
{\cal H}_{m-el} = \sum_m\ B_m \ e_m\  Q_m,
\end{equation}
where the $e$ variables are normalised strains, the $Q$ variables are $4f$ quadrupole operators and the $B_m$ are coupling coefficients, $m$ spanning the relevant symmetry allowed representations. In zero magnetic field, zero external strain and in the absence of any symmetry lowering transition, the mean equilibrium values $\langle e_m \rangle$ vanish. In the presence of a field, the $\langle e_m(H) \rangle$ become non-zero and describe the parastriction. The magneto-elastic coupling (\ref{hmel}) is included in the calculation of the magnetisation curves \cite{aleks,kleko} described in Section \ref{magh}. 

As mentioned in the introduction, there is experimental evidence for hybridisation of a phonon branch with a CEF transition in \tbti\ at very low temperature \cite{fennell2,guitt}. The whole vibronic problem, with mixed phonon-electron wave-functions, should thus be treated by expanding in (\ref{hmel}) the strain variables in terms of phonon operators. However, in this work, we approximate the vibronic problem by a dynamic Jahn-Teller (JT) effect, i.e. we introduce a dynamic distortion $\langle e \rangle_t$ along the equiprobable tetragonal cubic axes [100], [010] or [001], hence conserving overall cubic symmetry. A Hamiltonian of type (\ref{hmel}) is added to the CEF interaction, defined by ${\cal H}_{JT}^\alpha= B \langle e \rangle_t Q_{\alpha\alpha} = D_Q\ Q_{\alpha\alpha}$, where $Q_{ij} = \frac{1}{2} [J_i J_j+J_j J_i]$ and $\alpha=X,Y,Z$ is one of the three cubic \{100\} directions. The associated Hamiltonians ${\cal H}_{JT}^\alpha$ are written in the local frame:
\begin{eqnarray*}\label{dist}
{\cal H}_{JT}^X & = & \frac{D_Q}{3} \ [\frac{1}{2}\ Q_{xx}+\frac{3}{2}\ Q_{yy}+Q_{zz}-\sqrt{3} \ Q_{xy}\\ & & \hspace{15pt} -\sqrt{2} \ Q_{xz} +\sqrt{6} \ Q_{yz}] \\
{\cal H}_{JT}^Y & = & \frac{D_Q}{3} \ [\frac{1}{2}\ Q_{xx}+\frac{3}{2}\ Q_{yy}+Q_{zz}+\sqrt{3} \ Q_{xy}\\ & & \hspace{15pt} -\sqrt{2} \ Q_{xz} -\sqrt{6} \ Q_{yz}] \\
{\cal H}_{JT}^Z  & = & \frac{D_Q}{3} \ [2\ Q_{xx}+Q_{zz}+2\sqrt{2}\ Q_{xz}].
\end{eqnarray*}
It is crucial to note that, contrary to the magnetic moment operators $J_i$, the quadrupole moment operators $Q_{ij}$ generally couple the two states of the ground non-Kramers doublet: $\langle \psi_1 \vert Q_{ij} \vert \psi_2 \rangle \neq 0$, so that the dynamics of the system is deeply modified. For a given direction of the JT axis, the degeneracy of the ground doublet is lifted, and the new eigen-functions are close to the symmetric $\vert \psi_s \rangle$ and antisymmetric $\vert \psi_a \rangle$ combinations of the trigonal wave-functions $\vert \psi_1 \rangle$ and $\vert \psi_2 \rangle$. The ground state depends on the sign of the $D_Q$ parameter, whose value $D_Q=0.25$\,K was derived in our previous work \cite{bonville} from the energy of the lowest inelastic excitation. Then, the ground singlet is $\vert \psi_a \rangle = \frac{1}{\sqrt{2}} (\vert \psi_1 \rangle - \vert \psi_2 \rangle)$. We note that the single ion vibronic (or Jahn-Teller) coupling can lead to intersite quadrupole interactions, which could also play a role in defining the low temperature state of \tbti. 

The nearest neighbour exchange interaction is written in $J-J$ coupling, with the convention that a negative exchange integral corresponds to an AF coupling:
\begin{equation}\label{hech}
{\cal H}_{ex} =  -\sum_{<ij>}\ {\bf J}_i\ \tilde {\cal J}\ {\bf J}_j,
\end{equation}
where $\sum_{<ij>}$ means a summation over the first neighbour pairs.  The exchange tensor $\tilde{\cal J}$ is chosen to be anisotropic, with its symmetric part diagonal in a frame linked to a R-R bond \cite{malkin,bonverti}. For instance, the frame associated with the bond between site 1 (a/2,a/2,a/2) and site 2 (a/4,a/4,a/2) has unit vectors: {\bf a}$_{12}$=[001], {\bf b}$_{12}$=$\frac{1}{\sqrt{2}}[1\bar 10]$, {\bf c}$_{12}$ = $\frac{1}{\sqrt{2}}[110]$. The exchange tensor is identical in all these frames and is written:
\begin{equation}\label{jbond}
{\cal J} = \left(
\begin{array}{ccc}
{\cal J}_a & 0 & \sqrt{2}\ J_{DM} \\
 0 & {\cal J}_b & 0 \\
 -\sqrt{2}\ J_{DM} & 0 & {\cal J}_c
\end{array}
\right).
\end{equation}
The ${\cal J}_a$, ${\cal J}_b$ and ${\cal J}_c$ components represent the symmetric part and ${\cal J}_{DM}$ the antisymmetric Dzyaloshinski-Moriya part of the exchange. We use the anisotropic exchange tensor derived previously for \tbti\ \cite{bonville}, slightly modified to better match the diffuse scattering maps (see Section \ref{diff}).

For computing the magnetisation curves for each field direction, we introduce the Zeeman Hamiltonian ${\cal H}_Z$ and we perform, for each site $i$ of a tetrahedron and for each direction $\alpha$ of the JT axis, the diagonalisation of the following Hamiltonian:
\begin{equation} \label{hammh}
{\cal H}^i = {\cal H}_{CEF} + {\cal H}_{m-el} + {\cal H}_{JT}^\alpha + {\cal H}_Z + [{\cal H}_{ex}+{\cal H}_{dip}]^i,
\end{equation}
where $[{\cal H}_{ex}+{\cal H}_{dip}]^i$ is the part of the first-neighbour exchange and infinite range dipole-dipole couplings relative to site $i$, which is treated in a mean field self-consistent way within the 4 sites of a tetrahedron. The dipole sums are evaluated using the Ewald summation method \cite{wang}. Hence we limit our calculation to the case where the Tb moments in each of the 4 fcc sublattices forming the pyrochlore lattice are collinear. One thus obtains the magnetic structure (and the magnetisation) for a {\bf k}=0 propagation vector only.

The diffuse neutron scattering cross section is computed in the paramagnetic phase in zero field using the 4-site RPA outlined in Ref.\onlinecite{kao}. The starting points are the ${\cal H}_{CEF} + {\cal H}_{JT}^\alpha$ Hamiltonian and the Fourier transforms of the exchange and of the dipole-dipole interactions, the latter being obtained following Ref.\onlinecite{delmaes}. The components of the complex dynamic susceptibility $\chi({\bf Q},\omega)$ are the solutions of a set of linear equations solved using LAPACK routines. For a given {\bf Q} = {\bf q}+{\bf G}, where {\bf G} is a vector of the reciprocal space of the fcc lattice and {\bf q} belongs to the first Brillouin zone, the diffuse scattering intensity is obtained from the real part of the susceptibility as:
\begin{eqnarray}\label{chipr}
S_{el}({\bf Q},\omega) & \propto & \frac{\vert F({\bf Q}) \vert^2}{k_{\rm B}T}\ \sum_{\alpha,\beta,a,b}\ (\delta_{\alpha,\beta}-Q_\alpha Q_\beta) \cr
& \times & \exp[-i({\bf r}_a-{\bf r}_b).{\bf G}] \ {\rm Re} \ \chi^{\alpha,\beta}_{a,b}({\bf q},\omega),
\end{eqnarray}
where $F({\bf Q})$ is the magnetic form factor of the \tb\ ion, $\alpha$ and $\beta$ label spatial coordinates and $a$ and $b$ label the 4 sites on a tetrahedron. Since application of a magnetic field induces long range order in \tbti\ \cite{rule,sazonov,sazo2,guk1}, our RPA calculation of the diffuse scattering holds only in zero field, i.e. in the paramagnetic phase.

In our simulations of the present single crystal data, since an external axis is defined (the magnetic field or a particular plane of the reciprocal space), our assumption of a dynamic Jahn-Teller effect implies that an equal weight average over the three \{100\} directions of the JT axis must be performed. 

\begin{figure*}
\includegraphics[height=5.5cm]{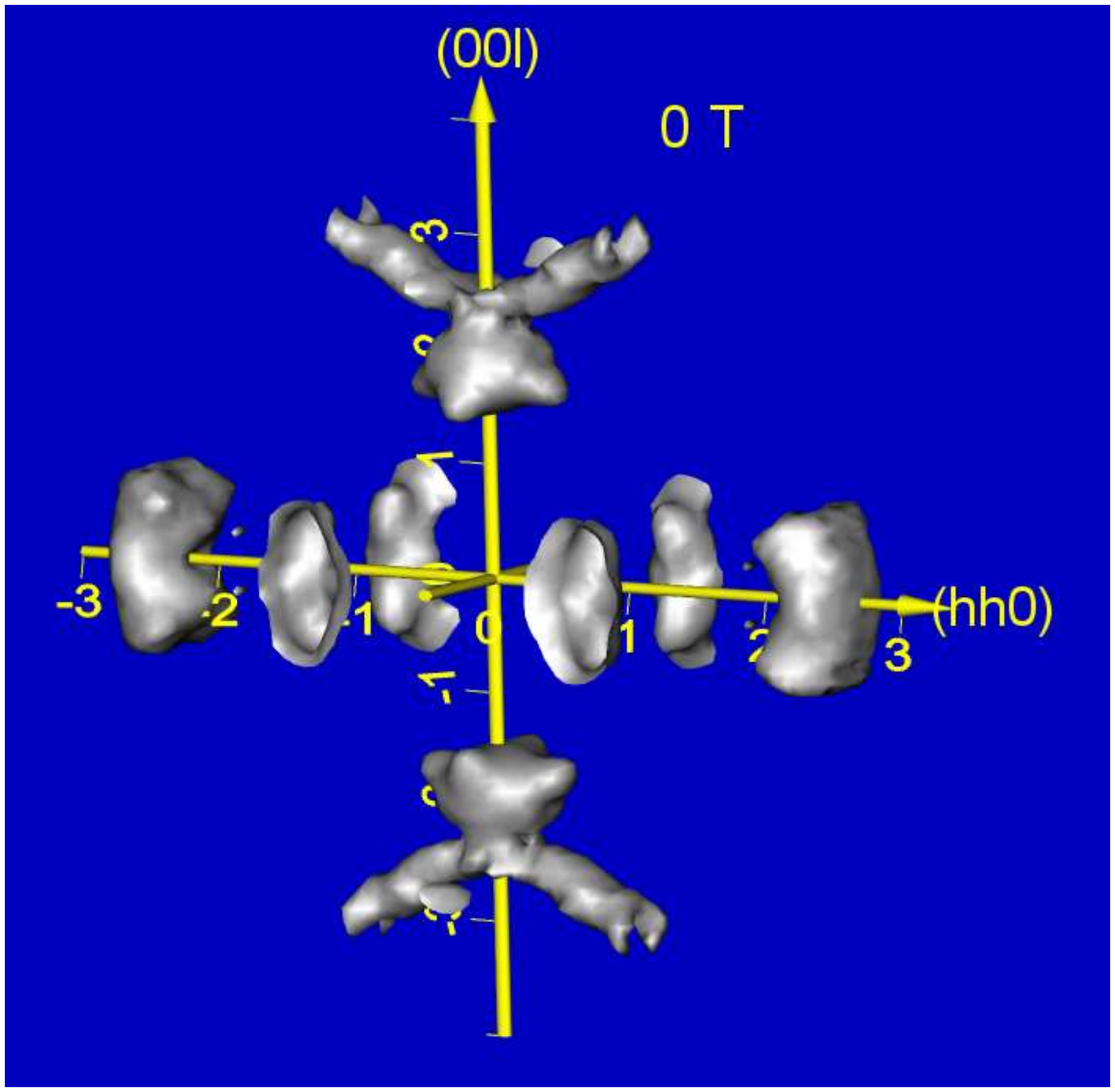}
\hspace*{13pt}
{\includegraphics[height=4.2cm]{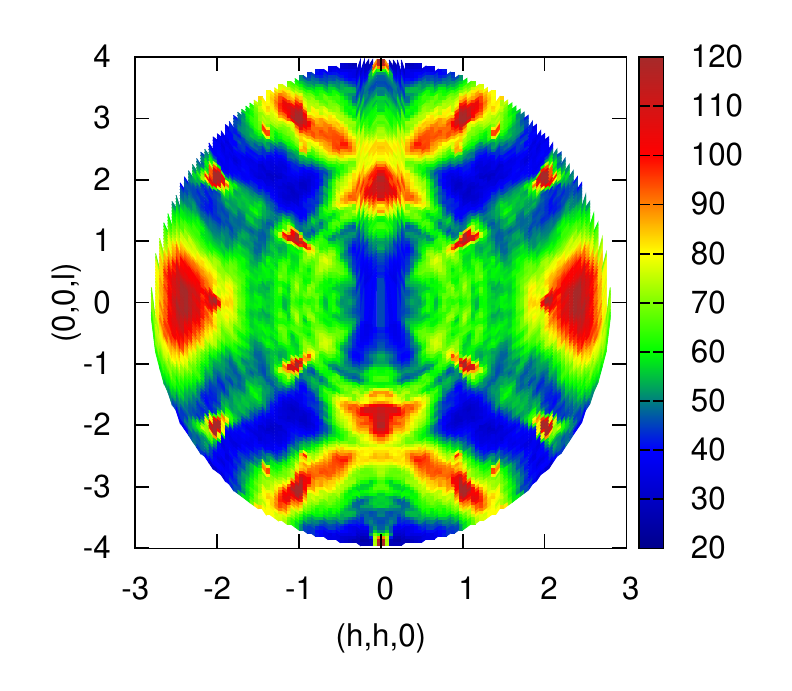}
\hspace*{5pt}
\includegraphics[height=3.8cm]{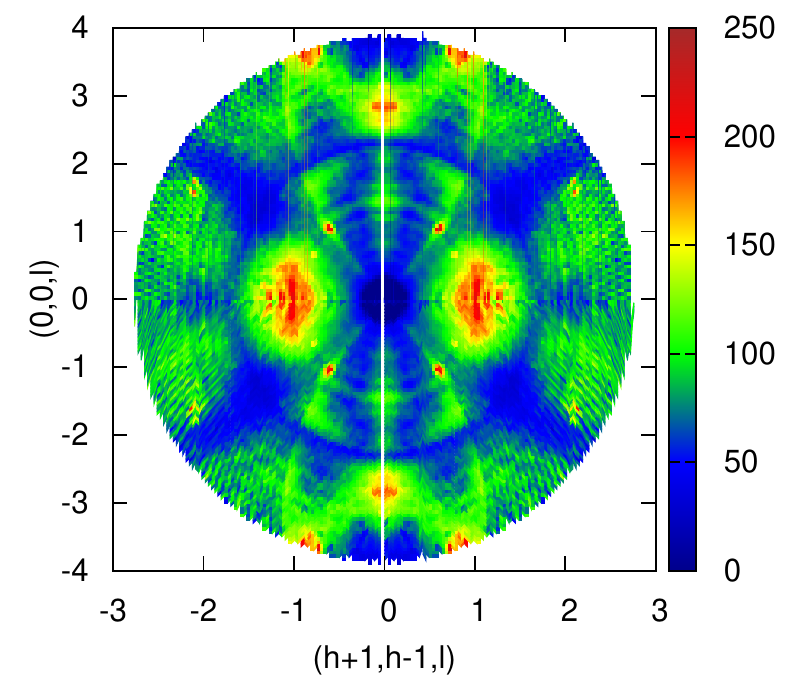}
}
\hspace*{120pt}
{\includegraphics[height=4.5cm]{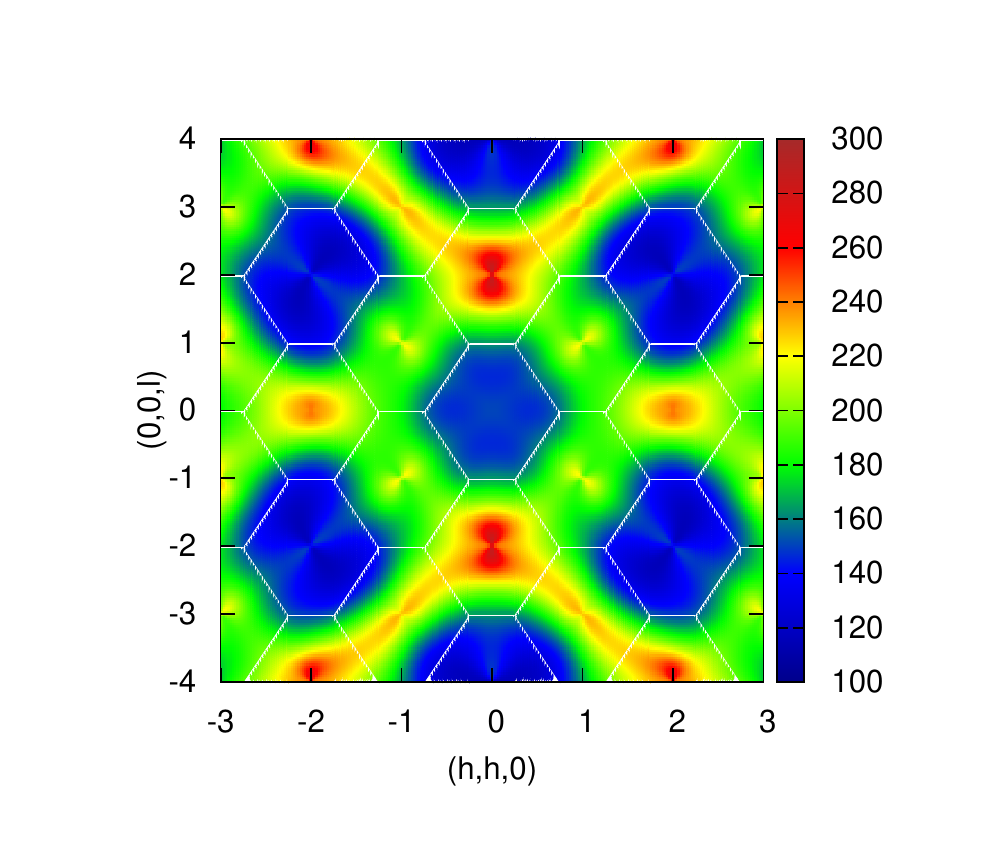}
\includegraphics[height=4.5cm]{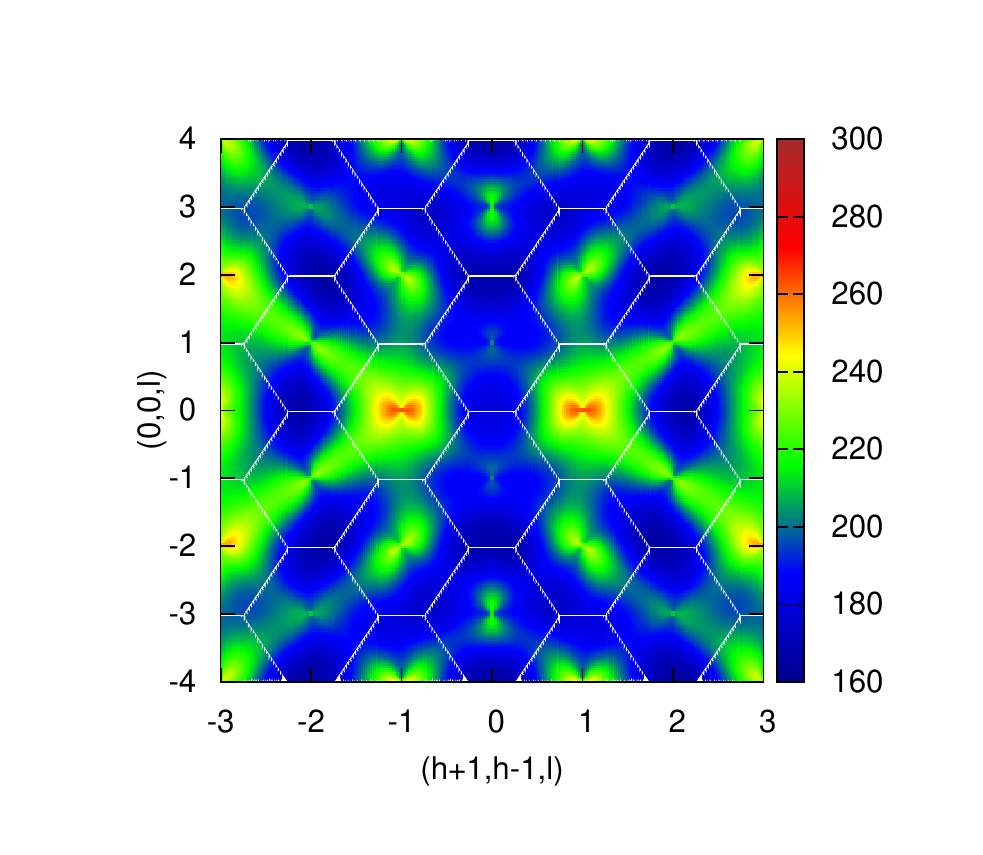}}
\caption{\label{diff160} Zero field neutron diffuse scattering maps in reciprocal space at 0.16\,K in \tbti: 3D equal intensity surface (left panel), experimental (upper central panel) and calculated (lower central panel) scattering in the (hhl) plane, experimental (right upper panel) and calculated (right lower panel) scattering in the (h+1,h$-$1,l) plane. The simulations were made in the presence of dynamic Jahn-Teller effect, with the anisotropic exchange tensor ${\cal J}_a=-$0.068\,K, ${\cal J}_b=-$0.196\,K, ${\cal J}_c=-$0.091\,K, ${\cal J}_{DM}$=0.}
\end{figure*}
\section{The diffuse neutron scattering at 0.16\,K}\label{diff}

Diffuse scattering maps were measured at 0.16\,K on the Super-6T2 diffractometer at the Orph\'ee reactor of the Laboratoire L\'eon Brillouin, Saclay (France) \cite{guk}, in zero magnetic field and with fields of 1 and 4\,T applied along [1$\bar1$0]. Data were collected using an area neutron detector ($\lambda$ = 2.35\,\AA) covering a 26$^\circ$ x 26$^\circ$ angular region, by rotating the sample about the [1$\bar1$0] axis with 0.1$^\circ$ step and recording a scattering pattern for two detector positions at 2$\theta$=17$^\circ$ and 40$^\circ$. In contrast to earlier diffuse scattering studies in \tbti\ \cite{petit,fennell}, the use of the area detector allows one to explore a large three-dimensional (3D) segment of the reciprocal space by transforming a complete set of area detector images in reciprocal space. The reconstructed volume was completed with its symmetrically equivalent orientations employing the Laue symmetry of the structure. In order to strengthen the contrast of the diffuse scattering images,
the regions of the nuclear and field induced magnetic Bragg reflections were excluded after the reconstruction. A constant background signal was subtracted.
\begin{figure*}
\includegraphics[height=5cm]{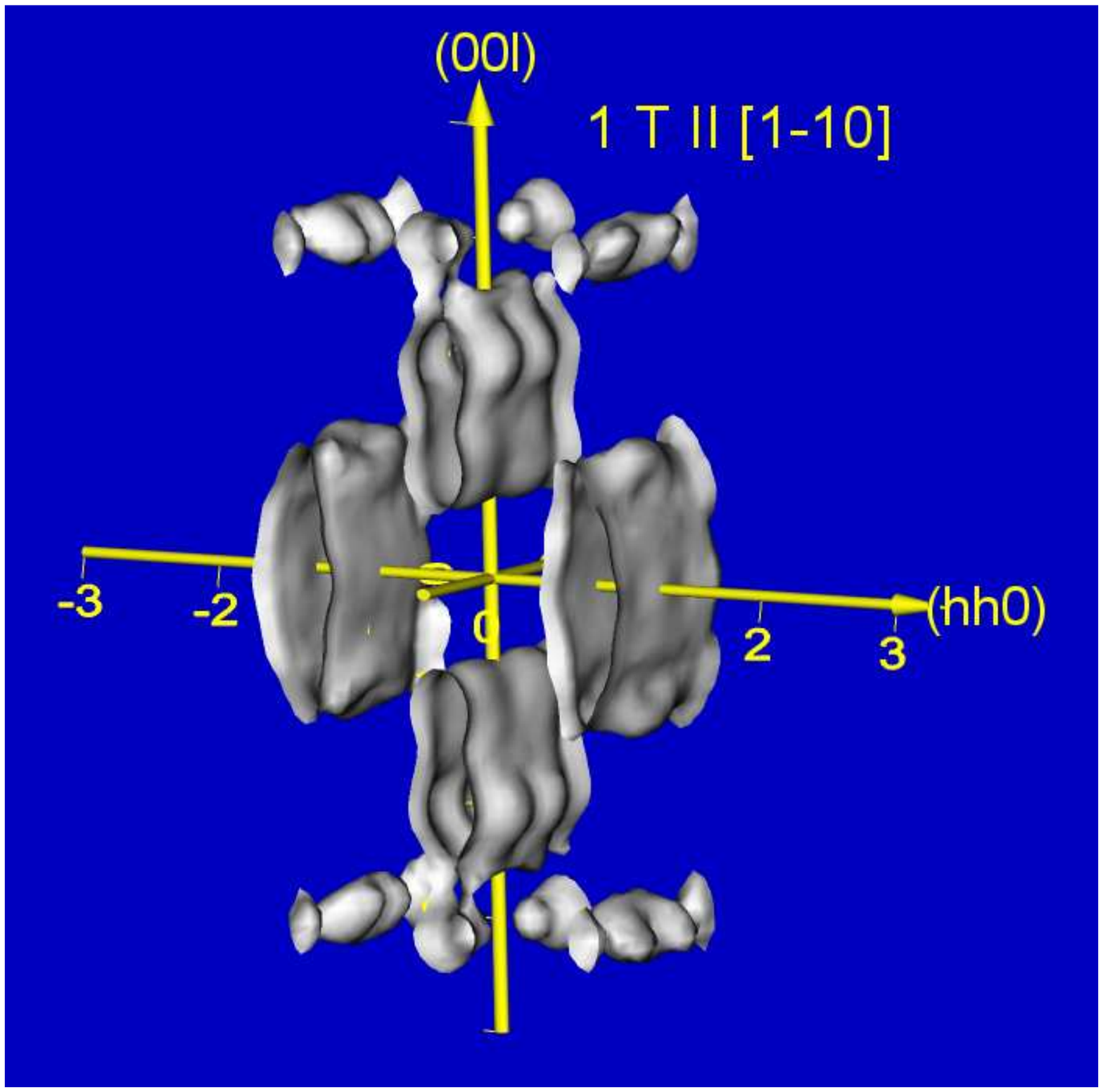}
\hspace*{15pt}
\includegraphics[height=5cm]{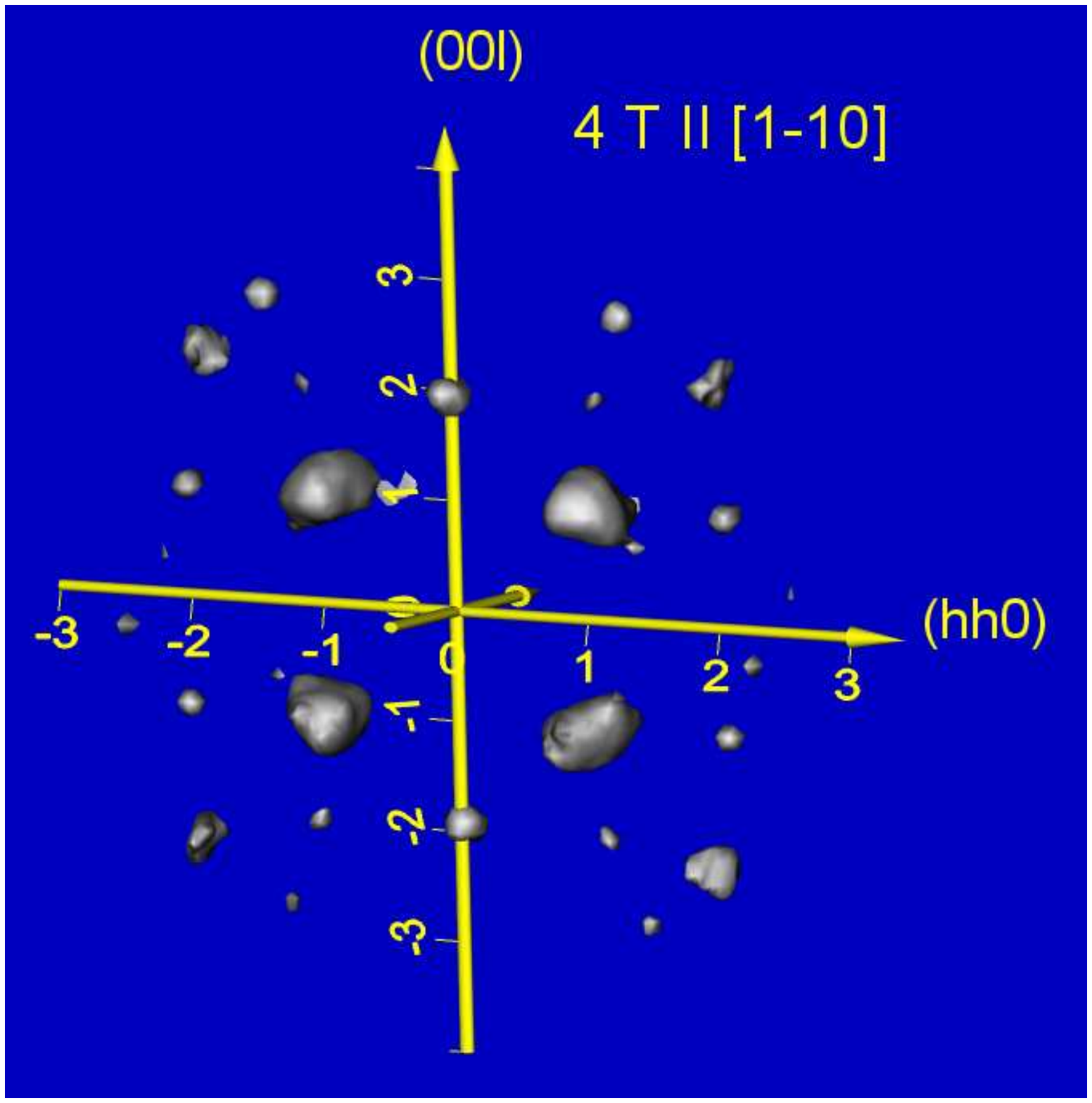}
\vspace*{-30pt}
\center{
\includegraphics[height=4cm]{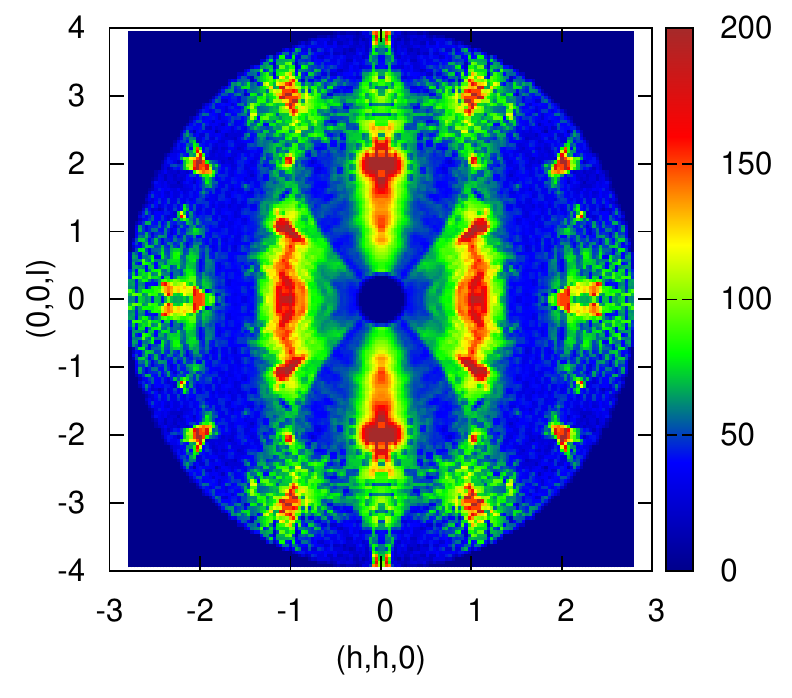}
\includegraphics[height=4cm]{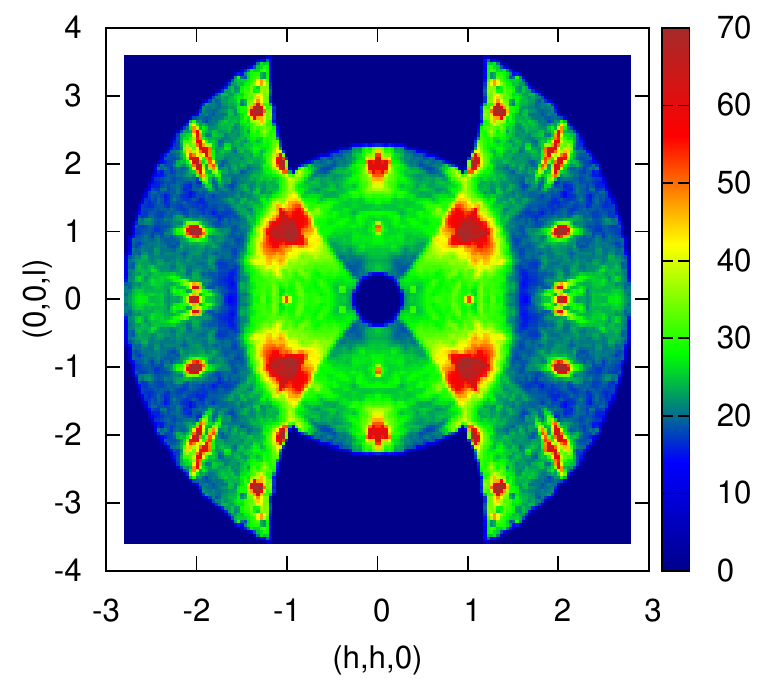}
}
\caption{\label{diffh160} Neutron diffuse scattering maps in reciprocal space at 0.16\,K in \tbti\ with a field applied along [1$\bar 1$0]. Upper panel: 3D equal intensity surfaces for a magnetic field of 1\,T (left) and 4\,T (right). Lower panel: cuts in the (hhl) plane of the maps at 1\,T (left) and 4\,T (right).}
\end{figure*}
A 3D representation of an equal scattering intensity surface in zero field is shown in the left panel of Fig.\ref{diff160}. From this pattern, the conventional 2D cuts (h,h,l) and (h+1,h$-$1,l), perpendicular to the [1$\bar1$0] axis, were obtained (upper central and right panels of Fig.\ref{diff160}). The (hhl) cut is in reasonable agreement with the diffuse scattering data using polarised neutrons of Ref.\onlinecite{fennell}. It shows asymmetric ``butterfly-like'' structures at (002) and (00$\bar 2$) and triangular spots near (220) and ($\bar2\bar 2$0), with strong intensity, and small pinch points at (111) etc. The intensity at the zone center is weak. The (h+1,h$-$1,l) cut shows strong intensity spots at (200) and weak intensity ``butterfly-like'' structures at (1$\bar 1$3).

The maps were simulated in the presence of dynamic Jahn-Teller effect using the diagonal exchange tensor elements (in K):
\begin{equation}\label{val}
{\cal J}_a=-0.068,\  {\cal J}_b=-0.196\  {\rm and}\  {\cal J}_c=-0.091,
\end{equation}
as previously determined in Ref.\onlinecite{bonville} (the absolute value of ${\cal J}_c$ is 8\% smaller that in Ref.\onlinecite{bonville}). The Dzyaloshinski-Moriya term ${\cal J}_{DM}$ is taken to be zero. The simulated maps, shown in the lower panels of Fig.\ref{diff160}, are seen to capture the main features of the experimental data, especially the strong intensity butterfly-like structures at (002) etc. in the (hhl) plane, and the very low scattering ``corridor'' along (1,$\bar 1$,l) and high intensity spots at (200) and (0$\bar 2$0) in the (h+1,h$-$1,l) plane.

Application of  magnetic field in  the [1$\bar1$0] direction at 0.16\,K induces magnetic order with two propagation vectors {\bf k}=0 and {\bf k}=[001], which leads to a strong decrease of diffuse scattering. Since a detailed description of the in-field AF structure has been given in Ref.\onlinecite{sazonov}, we concentrate here exclusively on the diffuse scattering results. Fig.\ref{diffh160} shows in-field diffuse scattering maps with 1 and 4\,T applied along [1$\bar1$0], the 3D equal intensity surfaces in the top panels and the (hhl) cuts in the lower panels. For these field values, our RPA approximation scheme, holding only in the paramagnetic phase, cannot be applied, so we shall limit ourselves  to a qualitative description of the phenomena. We recall that the field-induced structure involves the so-called $\alpha$ and  $\beta$ chains along two perpendicular directions. The $\alpha$ chains running along {\bf H} $\parallel$ [1$\bar1$0] have their local anisotropy axis at 36$^\circ$ from the field whereas the $\beta$ ones along [$110$] have their easy axis perpendicular to the field. In first approximation, the {\bf k}=0 structure is related to the ordering of the $\alpha$ chains and the {\bf k}=[001] structure to the ordering of the $\beta$ ones. In a field of 1\,T, only the long range ordered {\bf k}=0 structure is stabilized; the Tb moments in the $\alpha$ chains are easily aligned along their local axes, which leads to a strong increase of the intensities of the structurally allowed Bragg reflections, accompanied by a  decrease of diffuse scattering. The ``butterfly-like'' patterns are replaced by a disk shaped diffuse scattering concentrated at the forbidden reflections corresponding to the {\bf k} = [001] structure, like (110), (112) and (001). This remaining diffuse scattering  is due to the 1D short range  order of the $\beta$ chains. When the field is further increased to 4\,T, a 3D ordering of the $\beta$ chains occurs giving rise to the appearance of the forbidden reflections which violate the extinction rules of the fcc lattice \cite{sazonov}, and to the almost complete disappearance of the disk shape scattering. Diffuse scattering is however observed around the (111) reflection positions, possibly due to a small misorientation of the sample induced by the high applied field.

\section{Very low temperature isothermal magnetisation in \tbti}\label{magh}

In the spin liquid \tbti, application of a magnetic field induces magnetic order with {\bf k}=0 for {\bf H} // [110] \cite{sazonov} and {\bf H} // [111] \cite{sazo2}, and probably also for {\bf H} // [001] although no neutron diffraction data are available for this field direction. Actually, for {\bf H} // [110], an AF structure with {\bf k} = [100] coexists with the {\bf k}=0 structure above 2\,T and below 1\,K \cite{sazonov}, but it is not expected to contribute to the magnetisation. Therefore, the calculation sketched in Section \ref{rpa} can be applied to \tbti\ with a magnetic field applied along [110], [111] and probably [001].
\begin{figure*}
\vspace*{-12cm}
\includegraphics[height=20cm]{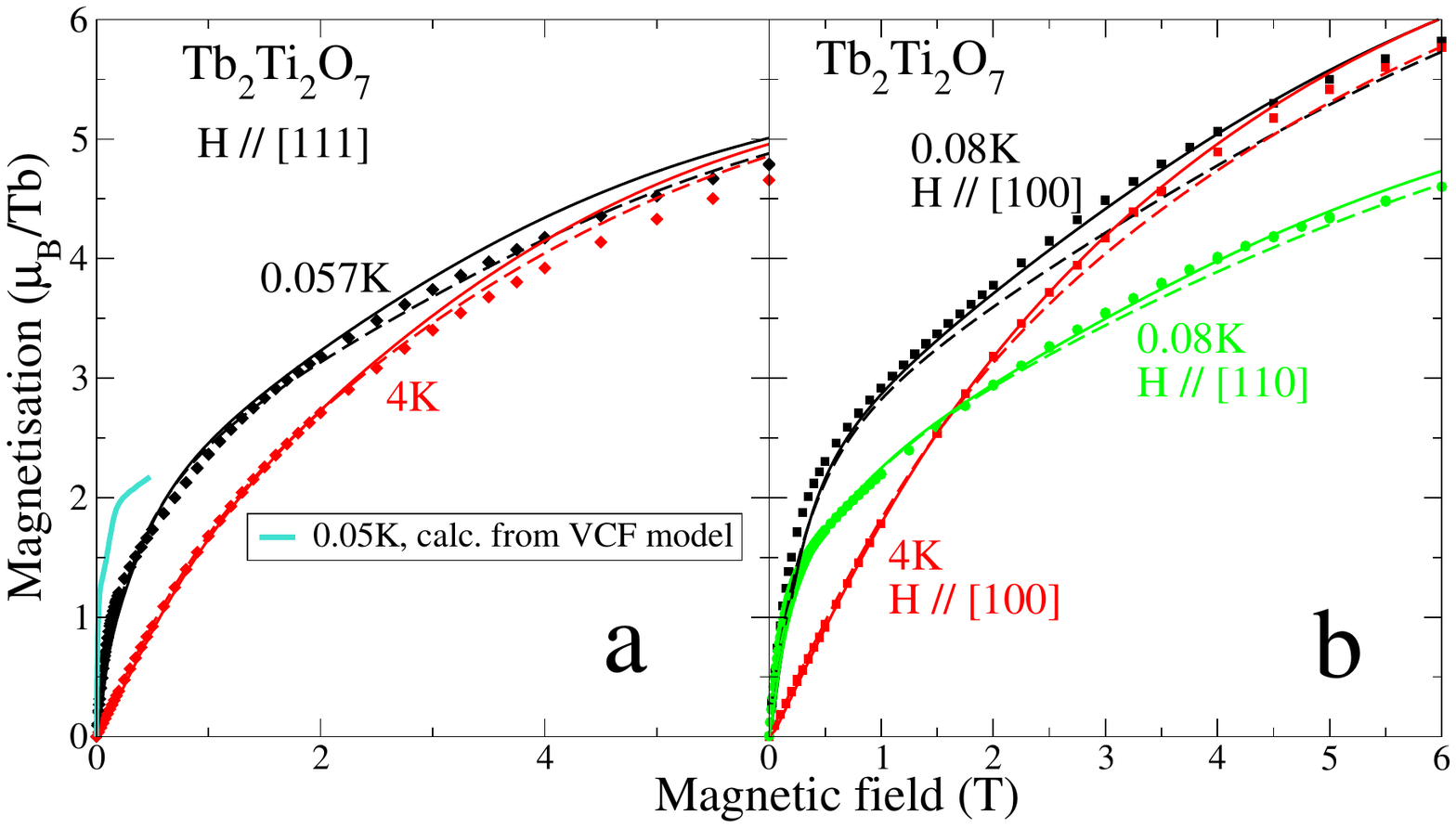}
\caption{\label{mh3axe} Isothermal magnetisation curves in \tbti: {\bf a} for {\bf H} // [111] at 0.057\,K and 4\,K; {\bf b} for {\bf H} // [100] and [110] at 0.08\,K and 4\,K. The experimental data for {\bf H} // [111] and [110] are taken from Ref.\onlinecite{lhotel}, those for {\bf H} // [100] from Ref.\onlinecite{lhotel1}. The calculated curves were obtained using the model described in the text with two assumptions: no magneto-elastic effects (dashed lines) and including the magneto-elastic interaction (solid lines), except for the solid blue line in {\bf a} which reproduces the prediction of the VCF model at 0.05\,K \cite{molavianging}, up to the highest field of the calculation (0.5\,T).}
\end{figure*}
Isothermal magnetisation curves in \tbti\ were measured in the 0.05\,K range in Refs.\onlinecite{lhotel} and \onlinecite{dunsiger}; they show a monotonic increase as the field increases (solid symbols in Fig.\ref{mh3axe}). For {\bf H} // [111], no magnetisation ``plateau'' predicted from the VCF model \cite{molavianging} and akin to that observed in spin ices for this same field direction \cite{matsu} is observed down to 0.05\,K, and we see no precursor effect of it. 
We recall however that, according to Ref.\onlinecite{molavianging}, a plateau is expected to be clearly visible at a temperature (0.02\,K) lower than that of these experiments ($\sim$0.050\,K).

For the calculation of the magnetisation curves, we use the values of the anisotropic exchange tensor as in the previous section. We have also taken into account the standard magneto-elastic (ME) interaction (\ref{hmel}), which yields giant magnetostriction effects \cite{aleks,kleko}. We limited ourselves to ME terms quadratic in the total angular momentum, following the formalism of Ref.\onlinecite{kleko} and using the ME parameters values derived therein. We have computed the 0.08\,K and 4\,K magnetisation curves for fields along [100] and [1$\bar1$0], and the 0.057\,K and 4\,K curves for {\bf H} // [111] up to 6\,T. At 4\,K, we set $D_Q$=0, since at this temperature the off-diagonal crystal field probably has a quite small effect, if any.

The comparison between experimental data and our calculation is shown in Fig.\ref{mh3axe}. The overall agreement is reasonably good both at very low temperature and at 4\,K. In order to assess the importance of the ME interactions, we have represented the calculated curves obtained without (dashed lines) and with (solid lines) ME effects. Inclusion of the ME interaction slightly modifies the magnetisation, especially at high fields. It yields a small enhancement, which results in a better agreement with experiment for {\bf H} // [100] and [1$\bar1$0] (Fig.\ref{mh3axe} {\bf b}), but not for {\bf H} // [111] (Fig.\ref{mh3axe} {\bf a}). For this latter field direction, we have also reproduced the curve calculated using the VCF model at 0.05\,K (blue line), taken from Ref.\onlinecite{molavianging}. Despite the limited field range, it is clear that it does not reproduce the data, which casts a doubt about the validity the VCF model for \tbti, at least as far as the magnetisation is concerned. Furthermore, this model predicts a sizeable variation of the shape of the low field magnetisation curve between 0.02\,K and 0.1\,K, with the appearance of a clear plateau at the lowest temperatures (0.02\,K). By contrast, the experimental data \cite{lhotel,dunsiger} (in agreement with our calculations) show that the shape of the magnetisation curve does not appreciably change between 0.05\,K and 0.3\,K. We believe the absence in \tbti\ of the magnetisation plateau expected for Ising spins is due to the vibronic nature of the ground state, which likely destroys the Ising character at the rare earth site linked with the bare trigonal crystal field eigen-functions. We note that these measurements of the isothermal magnetisation curves provide a {\it direct} test of the relevance of the VCF prediction. In contrast, the recent experimental data at very low temperature \cite{fritsch,yin} presented in favor of the VCF model are only an {\it indirect} evidence. They do not check the existence of the magnetization plateau in \tbti\ but probe low temperature anomalies which could have another origin.  

Although our model correctly reproduces the overall magnetisation behaviour and its anisotropy, some deviation occurs at the lowest temperature at low field, around 1\,T and below, mainly for {\bf H} // [111] and [100]. The curvature of the magnetisation as the field increases is not exactly reproduced. At higher fields, above 3-4\,T, the calculated points lie somewhat above the data points, especially for {\bf H} // [111] both at 0.08\,K and 4\,K. Keeping in mind that the uncertainty for such magnetic measurements is usually estimated to amount to a few percent, we can envisage various causes for these deviations. First, our model would not capture all the details of the field-induced magnetic structure, especially at low fields. Second, the effect of a slight field misalignment with respect to the crystal axes can also play a role. For {\bf H} // [1$\bar1$0], indeed, there occurs a ``spin melting'' near 1\,T where the two moments lying on sites with their ternary axis perpendicular to the field ($\beta$ sites) vanish \cite{sazonov}. The occurrence of the ``spin melting'' and the configuration of the $\beta$ moments are very sensitive to the alignment of the field with respect to the crystal axis. Finally, we note that the anisotropy of the magnetisation curves is mainly a trigonal crystal field effect; for fields above $\sim$2\,T, the Jahn-Teller interaction and its associated dynamic distortion play only a minor role. So, at low fields, calculation of the magnetisation curves can suffer from the fact that the dynamic Jahn-Teller model is only a first approximation to describe the ground state of \tbti.

\begin{figure*}
\vspace*{-11cm}
\includegraphics[height=20cm]{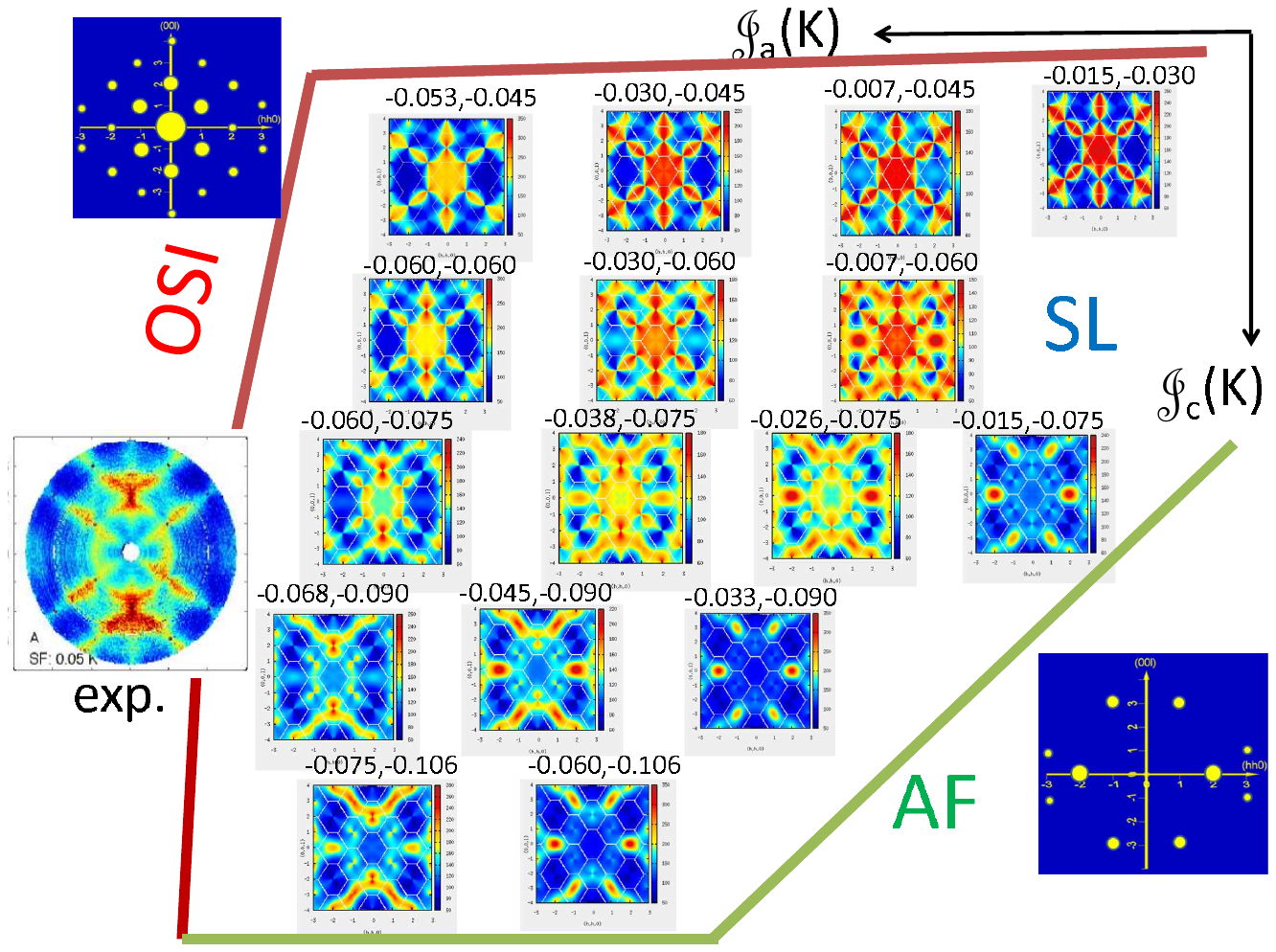}
\caption{\label{sfmap} Calculated diffuse scattering maps in the (hhl) plane of the reciprocal space for the Spin-Flip channel at 0.05\,K, according to the geometrical setup of Ref.\onlinecite{fennell}. The {\bf q}-maps are represented in the spin liquid (SL) phase of our model (see Ref.\onlinecite{bonville}), which stands as a wedge between the antiferromagnetic (AF) phase and the ordered spin ice (OSI) phase. The figure is a sketch of a cut in the exchange parameter phase space for ${\cal J}_b=-$0.196\,K \cite{note}; the numbers above each map are the values (in K) of ${\cal J}_a$ and ${\cal J}_c$. The map on the left labelled ``exp.'' is the experimental Spin-Flip diffuse scattering in \tbti\ at 0.05\,K from Ref.\onlinecite{fennell}; it has been placed close to the bottom left corner of the phase space for sake of comparison with the simulated maps. The maps with blue background placed in the AF and OSI phase represent the Bragg spots in the (hhl) plane for respectively the ``all in / all out'' and the ``two in - two out'' spin structures. }
\end{figure*}
\section {Discussion}

The Virtual Crystal Field approach of Ref.\onlinecite{molavian}, which is a first order perturbation theory, does not seem to explain the ground state of \tbti\ as observed experimentally. It seems that a more profound change of the ground state wave function (zeroth order perturbation) is necessary, and the dynamic JT interaction developed here is a first attempt to introduce such a quantum mixing. Should then \tbti\ be dubbed a ``Quantum Spin Ice'', as suggested by several authors? Although this term was initially proposed for \tbti\ in the context of the VCF approach \cite{molavian}, it is now more widely used when quantum fluctuations occur within a set of spin ice states \cite{benton,lee,gingclar}. Indeed \tbti, where \tb\ is a non Kramers ion, which shows dispersive excitations in a disordered ground state \cite{guitt} and where the likely presence of a vibronic coupling can induce an interaction between quadrupolar moments \cite{gehring}, is a good candidate. However, the small energy gap to the first excited CEF level renders the situation more complicated to treat theoretically than for the generic quantum spin ices with effective spin $\frac{1}{2}$ considered up to now. 

The simple distortion or Jahn-Teller model for \tbti\ implies that, for a certain range of exchange parameters, the mean field single ion ground state is a singlet, i.e. it is non-magnetic and thus could explain the absence of magnetic ordering down to the lowest temperature. However, energy resolved neutron scattering shows that elastic or quasi-elastic scattering is dominant at very low temperature \cite{taniguchi,fritsch,petit,fennell,takatsu}, meaning that the ground state has a non-zero magnetic moment which is static at the time scale of the experiment.

The fact that our model can reasonably well reproduce the {\bf q}-maps in reciprocal space of the inelastic \cite{petit} and diffuse (i.e. energy integrated) scattering (Ref.\onlinecite{petit} for the Spin-Flip (SF) and Non-Spin-Flip scattering and present work for the total scattering) probably shows that it correctly represents the spin correlations. However, it situates their characteristic energy in the inelastic channel rather than in the elastic one, since a non-magnetic ground state does not give rise to elastic scattering. In terms of the mixed eigen-states $\vert \psi_a \rangle$ and $\vert \psi_s \rangle$ introduced in Section \ref{rpa}, one can define a ``transition vector'' {\bf T} = $\langle \psi_a \vert {\bf J} \vert \psi_s \rangle$, whose square modulus is the intensity of the low energy inelastic mode. The {\bf T} vector is directed along the local trigonal axis since its only non-zero component is $T_z = \langle \psi_a \vert J_z \vert \psi_s \rangle=\langle \psi_1 \vert J_z \vert \psi_1 \rangle$, and it is solely defined  by the CEF wavefunctions. 
So the intensity of the calculated diffuse scattering involves the pure ``inelastic'' CEF quantity $\| {\bf T} \|^2$, the subsequent energy integration yielding expression (\ref{chipr}), proportional to ${\rm Re}\ \chi$ according to the Kramers-Kronig theorem, and its {\bf q}-dependence is mainly due to the anisotropy of the exchange tensor. 

To emphasize the difference of the diffuse scattering in \tbti\ with respect to those in pyrochlore systems with dominant spin-ice or antiferromagnetic spin correlations, we have computed the Spin Flip scattering maps in the (hhl) plane within the spin-liquid phase of our model \cite{bonville} (see Figure \ref{sfmap}). Near the upper border of the SL phase with the Ordered Spin Ice (OSI) phase (${\cal J}_c = -0.030, -0.045$\,K), the SF maps are very close to those of a spin ice \cite{bramwell}, where the spin correlations are of the type ``two in - two out''. The corresponding Bragg spots (map with blue background in the OSI phase) are located at the pinch points of the SF map. Near the border with the antiferromagnetic (AF) phase, the SF scattering consists only in broadened peaks at positions (220), (113), etc. These positions are those of the Bragg peaks of the ``all in / all out'' spin structure, as shown in the map with blue background in the AF phase. The sketch of the SL phase in Fig.\ref{sfmap} allows one to follow the progressive transformation of the SF maps as one scans the \{${\cal J}_a,{\cal J}_c$\} plane. It shows that the SF map is very specific to a given set of exchange integrals. The map closest to the experimental data (${\cal J}_a=-$0.068\,K, ${\cal J}_c=-$0.090\,K), near the bottom left corner, is different from the two above described limiting cases and must therefore represent a special type of spin correlations occuring in \tbti, probably consisting in a mixture of spin-ice and AF correlations. Within our model, \tbti\ lies very close to the border with the long range ordered OSI phase. This could explain the presence of spin ice like features in the AF spin correlations \cite{fennell,petit,fritsch,guitt,fennell2} and the stabilisation of the OSI phase in the sibling material \tbsn\ below 0.87\,K \cite{mirb05}.  

In this latter compound, a nuclear Schottky anomaly is present at low temperature and the hyperfine field derived from it is 180\,T, corresponding to an effective magnetic moment of 4.5\,\mub/\tb\ \cite{note1}, using the $^{159}$Tb$^{3+}$ hyperfine constant of 40(4)\,T/\mub\ \cite{dunlap}. This effective moment of 4.5\,\mub\ is reduced with respect to the spontaneous moment 5.9(1)\,\mub\ measured by neutron diffraction, which is attributed in Ref.\onlinecite{mirb05} to persisting spin fluctuations in the OSI phase. More precisely, the hyperfine Schottky anomaly is depleted in case the electronic spin fluctuation frequency $\nu_{4f}$ is not much slower than the nuclear relaxation frequency 1/$T_1$ \cite{bertin}, i.e. if $\nu_{4f} \sim 1/T_1$. The presence of a similar nuclear Schottky anomaly in the very low temperature specific heat of \tbti, with a hyperfine field of 142\,T \cite{yaouanc}, is rather unexpected in a material with no magnetic ordering. It shows that there is a hyperfine field at the nucleus site, linked in this case with the short range dynamically correlated \tb\ magnetic moments. The moment value derived from the 142\,T hyperfine field is 3.6(4)\,\mub. 
Thus, from the point of view of the nuclear specific heat, \tbsn\ and \tbti\ behave very similarly, although the former orders magnetically while the latter does not. As a further analogy, in order to account for the inelastic neutron scattering maps in the OSI phase of \tbsn\ \cite{petit_tbsn}, it is necessary to introduce an off-diagonal crystal field term which was taken as a tetragonal distortion. The $\mu$SR data in \tbti\ \cite{gardner} are also very similar to those in \tbsn\ \cite{dalm_tbsn,bert}, revealing sizeable fluctuating moments in both compounds down to the lowest temperature.

Inspection of the bulk of experimental data in \tbti\ at very low temperature suggests the following: on the one hand, experiments probing the behaviour of \tbti\ with a small characteristic time $\tau_c$, like the quasi-elastic neutron scattering ($\tau_c \sim \hbar/\delta E$, where $\delta E$ is the energy resolution, i.e. $\tau_c \simeq 10^{-11}$\,s for $\delta E=0.07$\,meV), the specific heat ($\tau_c \sim T_1$, where $T_1$ is the $^{159}$Tb nuclear relaxation time), $\mu$SR spectroscopy ($\tau_c \sim \tau_\mu$, where $\tau_\mu \simeq 2$\,$\mu$s is the muon lifetime), reveal the short time behaviour of the \tb\ ions. This behaviour is that of an ion submitted to a dynamic non-zero exchange/dipolar field, whose magnitude ($\sim$1\,T) is enough, in the presence of the JT or vibronic coupling, to polarise the moment during the lifetime of the spin correlations. On the other hand, experiments with a long characteristic time, like the magnetisation measurements ($\tau_c \sim \tau_m$, where $\tau_m \simeq 100$\,s is the magnetometer measuring time) or without well defined characteristic time, like the diffuse neutron scattering measurements, seem to be correctly described by the dynamic JT model, with its non-magnetic ground state. 

\section{Conclusion}

We have shown that it is possible to account for some quite different properties of the pyrochlore spin-liquid candidate \tbti\ at very low temperature, i.e. the neutron diffuse scattering and the isothermal magnetisation, by a model which approximates the probably complex vibronic ground state of this material by a dynamic Jahn-Teller coupling within the ground electronic doublet. The success of the model in reproducing the magnetisation curves is linked with the mixed character of the ``tunnel-like'' antisymmetric ground wave-function. Absence of this zero-order mixing automatically leads to a {\it kagom\'e}-ice state for intermediate field values along [111] and thus to a magnetisation plateau. As to the diffuse scattering, its calculated intensity reflects that of the transition between the mixed Jahn-Teller states, and the {\bf q}-dependence of the maps reflects the anisotropy of the exchange tensor. However, this approach does not provide a fully correct picture for \tbti\ since it predicts a non-magnetic ground state, in contradiction with the observation of a sizeable quasielastic neutron scattering and of a magnetic hyperfine Schottky anomaly in the specific heat. We believe it could anyway be a starting point while awaiting for a more elaborate treatment of the peculiar coupling between the phononic and electronic degrees of freedom which seems to be at play in \tbti\ at very low temperature.

\begin{acknowledgments}
We are indebted to B. Z. Malkin for his help with the magneto-elastic calculations. We are grateful to J. Robert and A.P. Sazonov for fruitful discussions, to C. Decorse and G. Dhalenne, and to P. Dalmas de R\'eotier and Ch. Marin, for providing the \tbti\ single crystals. We thank E. Lhotel and C. Paulsen for communicating unpublished results. 
\end{acknowledgments}


\begin{thebibliography}{99}
\bibitem{note} The phase diagram in Ref.\onlinecite{bonville} is represented in the space of the molecular constants $\lambda$. The relationship with the exchange constants used here is: ${\cal J}$=$(g_J \mu_{\rm B})^2/k_{\rm B}\ \lambda$, i.e. ${\cal J}$(K)=1.51 $\lambda$(T/$\mu_{\rm B}$).
\bibitem{note1} The hyperfine field and the effective moment values in Ref.\onlinecite{mirb05} have been erroneously quoted as 135\,T and 3.35\,\mub/\tb\ respectively. Their actual values in \tbsn\ are 180\,T and 4.5\,\mub/\tb\ respectively.
\end{thebibliography}
\end{document}